\documentclass[11pt,reqno]{article}
\usepackage{amsmath}
\usepackage{amsfonts}
\usepackage{amssymb}
\usepackage{hyperref}
\usepackage{latexsym}
\usepackage[dvips]{graphicx}
\usepackage{epsf}
\usepackage{color}

\allowdisplaybreaks \numberwithin{equation}{section}

\newcommand{\be}{\begin{equation}}
\newcommand{\ee}{\end{equation}}
\newcommand{\bea}{\begin{eqnarray}}
\newcommand{\eea}{\end{eqnarray}}

\newcommand{\benn}{\begin{equation*}}
\newcommand{\eenn}{\end{equation*}}

\newcommand{\f}{\frac}

\newcommand{\Tr}{{\rm Tr}}
\newcommand{\la}{\langle}
\newcommand{\ra}{\rangle}

\let\a=\alpha \let\b=\beta  \let\g=\gamma  \let\d=\delta
       \let\k=\kappa \let\l=\lambda
              \let\r=\rho \let\om=\omega
\let\s=\sigma \let\t=\tau     
    
 \let\D=\Delta   \let\L=\Lambda 
         
\let\Om=\Omega  \let\eps=\epsilon

\def\PPP{{\cal P}} 
 
\def\TT{{\cal T}} 
  \def\OO{{\cal O}}

\newcommand{\Nt}{\tilde{N}}

\newcommand{\Vt}{\tilde{V}}

\newcommand{\cL}{\mathcal{L}}

\newcommand{\cR}{\mathcal{R}}

\begin{document}

\thispagestyle{empty}
\begin{flushright} \small
AEI-2014-042
\end{flushright}
\bigskip

\begin{center}
 {\LARGE\bfseries   Spacetime condensation in (2+1)-dimensional CDT \\
 from a Ho\v{r}ava-Lifshitz minisuperspace model 
 }
\\[10mm]
{\large Dario Benedetti${}^a$ and Joe Henson${}^{b}$}
\\[3mm]
{\small\slshape
${}^a$ Max Planck Institute for Gravitational Physics (Albert Einstein Institute), \\
Am M\"{u}hlenberg 1, D-14476 Golm, Germany \\ 
{\rm Present address}: Laboratoire de Physique Th\'eorique, CNRS-UMR 8627,\\ Universit\'e Paris-Sud 11, 91405 Orsay Cedex, France\\
${}^b$ HH Wills Physics Laboratory, University of Bristol, Bristol BS8 1TL, United Kingdom\\
\vspace{.3cm}
 {\upshape\ttfamily dario.benedetti@th.u-psud.fr, j.henson@gmail.com} }
\end{center}
\vspace{5mm}

\hrule\bigskip

\centerline{\bfseries Abstract} \medskip
\noindent

A spacetime condensation phenomenon underlies the emergence of a macroscopic universe
in causal dynamical triangulations, where the time extension of the condensate is strictly smaller than the total time. 
It has been known for some time that the volumes of spatial slices in the bulk of the macroscopic universe follow a time evolution which resembles that of a sphere, and their effective dynamics is well described by a minisuperspace reduction of the general relativistic action.
More recently, it has been suggested that the same minusuperspace model can also provide an understanding of the condensation phenomenon itself, thus explaining the presence of an extended droplet of spacetime connected to a stalk of minimal spatial extension.
We show here that a minisuperspace model based on the general relativistic action fails in that respect for the (2+1)-dimensional case, while a successful condensation is obtained from a minisuperspace model of Ho\v rava-Lifshitz gravity.

\bigskip
\hrule\bigskip
\tableofcontents
\bigskip
\hrule\bigskip

\section{Introduction}

Defining a theory of quantum gravity is a notoriously difficult problem, which has given rise to many different ideas and approaches.
Within this context, the Causal Dynamical Triangulations (CDT) program \cite{Ambjorn:2012jv} stands out as an approach capable of producing many fully nonperturbative and background independent results.  Its central proposals are to seek a non-perturbative definition of the theory through the path integral, to define this via a piecewise-flat discretisation, to vary the connectivity between the simplex ``building blocks'' (rather than varying the geometrical properties of these blocks), and finally, to impose a ``causal'' restriction on the set of configurations to be summed over.  The addition of this restriction finally produced a model from which strong evidence for the recovery of a well-behaved extended spacetime could be derived in 3+1 D via computer simulations, and other promising results.

The causal restriction requires a preferred foliation (or ``time-slicing'') at the discrete level,\footnote{Although see \cite{Jordan:2013iaa} for evidence that the desirable properties of the model can be preserved without obviously invoking a foliation, but rather enforcing a more local version of causality.} violating the symmetries of GR.  The initial hope was that the desired symmetry would re-emerge if and when an appropriate continuum limit could be taken, and this has not yet been ruled out.  One possibility is that, equipped with the correct limiting procedure, CDT will provide a lattice implementation of the asymptotic safety scenario \cite{Weinberg:1980gg,Niedermaier:2006wt,Percacci:2007sz,Litim:2011cp,Reuter:2012id}.  More recently, strong connections have been found \cite{Horava:2009if,Ambjorn:2010hu} to Ho\v rava-Lifshitz (HL) gravity \cite{Horava:2008ih,Horava:2009uw}, which explicitly breaks foliation-independence, raising the possibility that this theory is the more natural continuum limit of CDT.  The possibility of finding multiple, physically different, continuum limits also remains open \cite{Ambjorn:2014gsa}.

In the simulations, three phases were found in the (3+1)-dimensional CDT phase diagram, only one of which gave rise to anything resembling an extended spacetime \cite{Ambjorn:2004qm,Ambjorn:2005xx}.  In one of the phases, sometimes refered to as the ``$A$'' phase, spatial volume is essentially spread randomly over all times in the simulation.  In the ``$B$'' phase all spatial volume is concentrated at one value of time, leaving minimal volume on all other slices.  In the well-behaved ``$C$'' phase,  the vast majority of the simplices are to be found in a ``droplet'' with non-trivial time extension, but outside of this contiguous region the spatial volume is near-minimal.  This spatially trivial region is referred to as the ``stalk''.    For the droplet region, computational results for spatial volume as a function of time (the ``volume profile'') can be compared to minisuperspace models derived from GR, with good results for both expectation values and fluctuations \cite{Ambjorn:2005xx}.

Bogacz, Burda and Waclaw \cite{Bogacz:2012sa} took this analysis one step further by studying a discrete path integral for the minisuperspace model.  They noted that this model is similar to the well-studied  ``balls in boxes'' statistical models, which exhibit the same kind of ``condensation'' behaviour observed in the CDT simulations.\footnote{Connections in quantum gravity between condensation phenomena and the recovery of extended spacetimes have also been investigated from the perspective of group field theory \cite{Gielen:2013naa,Gielen:2014uga}.}  They found that this simple model not only explains the volume profile of the droplet in phase $C$, but also the presence of a stalk, and the other two phases.  Furthermore the model possesses a phase diagram qualitatively matching the explored region of the CDT phase diagram, and suggesting possible new phases. This could be interpreted as further evidence for a connection between standard GR and CDT theory, but it would be interesting to explore this question further.  For example, it is of interest to ask whether the same connections hold between GR-inspired minisuperspace models and CDTs in 2+1 dimensions, and if not, to ask what kind of theory might reproduce the qualities of the phases seen there.

It is often convenient and interesting to consider lower-dimensional models, and quantum gravity in 2+1 dimensions has been a very active field of research over the years \cite{Carlip:1998uc}.
The same idea applies to CDT, which becomes so simple that it can be solved analytically in 1+1 dimensions \cite{Ambjorn:1998xu}. In comparison with this even lower-dimensional case, CDT theory in 2+1 dimensions has seen relatively few advances on the analytical front (\textit{e.g.}~\cite{Ambjorn:2001br,Benedetti:2007pp}), but it has proved to be a fruitful testbed for many ideas at the numerical level \cite{Ambjorn:2000dja,Ambjorn:2002nu,Benedetti:2009ge,Kommu:2011wd,Anderson:2011bj,Cooperman:2013mma,Budd:2011zm,Budd:2013waa,Jordan:2013iaa}, where it has been found that the 2+1D model possesses phases similar to phases $A$ and $C$ of the 3+1D model.
In particular, both the 1+1 \cite{Ambjorn:2013joa} and (2+1)-dimensional case \cite{Benedetti:2009ge,Anderson:2011bj,Budd:2011zm} have been useful in establishing tighter links to HL gravity.\footnote{Obviously the connections are to its lower-dimensional versions, which have been studied for example in  \cite{Horava:2008ih,Sotiriou:2011dr,Benedetti:2013pya}.}  Below, we add to these studies by showing that the analysis of Bogacz et al. can only be extended to the 2+1D case if the minisuperspace model employed is derived not from GR, but from HL gravity.

In Sec.~\ref{Sec:data} we present results from our numerical simulations of (2+1)-dimensional CDT. These are not the first simulations of this sort to have been carried out, but we review the more recent data here because they are obtained for a larger system size than previous simulations, and because the details of the results will be essential to the subsequent discussion.
In Sec.~\ref{Sec:BIB} we review the balls-in-boxes models and their relation to CDT. Besides recalling known facts, we also make some key observations concerning the absence of a Hamiltonian constraint in CDT, and show that the droplet condensation of (2+1)-dimensional CDT cannot be explained by the corresponding minisuperspace model based on general relativity.
In Sec.~\ref{Sec:miniHL} we introduce a minisuperspace reduction of a Ho\v rava-Lifshitz gravity model in 2+1 dimensions, which we propose as an effective model for CDT.
We study the minimisation of the action and show that a droplet condensation wins over a completely constant or a purely oscillatory solution.
Lastly, we compare the predictions of our model with the CDT data in Sec.~\ref{Sec:comparison}, where we also discuss the continuum limit, and conclude in Sec.~\ref{Sec:concl} with a discussion of open questions.

\section{Spatial volume dynamics in (2+1)-dimensional CDT}
\label{Sec:data}

To obtain a non-perturbative evaluation of a path integral in quantum field theory, it is standard to replace continuum spacetime with a fixed lattice.  This allows computer simulations to estimate quantities of interest, besides providing an approximate definition of an object that otherwise has no meaning.  Recovering the continuum theory is a delicate process that relies on the theory of critical phenomena and the renormalisation group.

The CDT approach follows these QFT techniques as closely as possible, but to deal with a theory of dynamical geometry, with no background spacetime fixed \textit{a priori}, the fixed lattice is replaced with an ensemble of random triangulations.
More specifically, each of these triangulations is a simplicial manifold,  i.e. a collection of $d$-dimensional flat simplices (the generalisation of triangles and tetrahedra) glued along their $(d-1)$-dimensional faces and such that the neighbourhood of any vertex is homeomorphic to a $d$-dimensional ball.  A dynamical triangulation is one in which all the simplices are taken to be equilateral, with edge length $a$.  In the simulations we usually work in the canonical ensemble of triangulations with a fixed number of $d$-simplices $N_d$, which we will denote $\TT$. The triangulations in this ensemble are obtained by ``gluing'' the $N_d$ simplices in all possible ways allowed by the simplicial manifold condition.

In older models of dynamical triangulations, where only the spacetime topology is fixed, the path integral is dominated by badly behaved configurations that do not resemble extended spacetimes. In CDT models a further restriction is imposed on the ensemble: only triangulations with a global time foliation, with respect to which no spatial topology change occurs, are allowed. For more details on the geometrical meaning of this restriction and on its implementation see \cite{Ambjorn:2001cv}.

The dynamics of the model is defined by the Euclideanised path integral, or partition function, which in the grand canonical ensemble is given by
\be \label{Z}
Z_{\rm gc} = \sum_{N_d} \sum_{\TT} \tfrac{1}{C(\TT)}\, e^{-S(\TT)} \, ,
\ee
where $S(\TT)$ is the bare action, and $C(\TT)$ is the order of the automorphism group of $\TT$, a symmetry factor naturally appearing when summing over unlabelled triangulations. A simple choice for the bare action is the Einstein-Hilbert action adapted to a simplicial manifold, known as the Regge action. When all edge-lengths are equal, the Regge action reduces to the convenient form
\be \label{action}
S(\TT) = \k_d N_d - \k_{d-2} N_{d-2} \, ,
\ee
where $\k_d$ and $\k_{d-2}$ are two coupling constants depending on the cosmological and Newton's constant appearing in the Regge action, and $N_{d-2}$ is the number of $(d-2)$-dimensional subsimplices (also called bones or hinges).
In general we will use $N_n$ to denote the number of $n$-dimensional simplices ($n=0,1,\ldots d$).
Because of the foliation in CDTs, it is possible to differentiate between spacelike and timelike edges\footnote{We are using here a Lorentzian language even though we have already carried out a Wick rotation and our signature is Euclidean; due to the preferred foliation, we can still identify the ``spacelike'' edges as the ones lying entirely within one slice of the foliation.}, and as a consequence we can count them separately as $N_1^s$ and $N_1^t$ respectively. Similarly $N_2^s$ is the number of triangles lying entirely within one slice of the foliation, $N_2^t$ is the number of those having one vertex on an adjacent slice, and so on. Finally we will write $N_{(m,d+1-m)}$ for the number of $d$-dimensional simplices having $m$ vertices on one slice and the remaining ones on an adjacent slice.
Despite the length of this list, due to topological relations \cite{Ambjorn:2001cv} \eqref{action} is the most general linear action that we can write with such variables for $d=3$, which is the case we are interested in.

In our simulations we will use these topological constraints to trade the variable $N_1$ for $N_0$, which is easier to keep track of, and replace \eqref{action} for $d=3$  by
\be \label{action-3d}
S(\TT) = \k_3 N_3 - \k_0 N_0 \, .
\ee
Furthermore, as we mentioned, in the computer simulations we work at fixed volume, and hence we replace \eqref{Z} by
\be \label{Z_N}
Z_{\rm c}(N_3) =  \sum_{\TT} \tfrac{1}{C(T_N)}\,e^{\k_0 N_0} \, ,
\ee
where we have made use of the simple form of the action \eqref{action-3d}. Note that the partition function $Z_{\rm c}$ is the discrete Laplace transform of $Z_{\rm gc}$ with respect to $N_d$.
The expectation value of an observable $A$ can be calculated from
\be
\la A \ra_{N_3} = \frac{1}{Z_{\rm c}(N_3)} \sum_{\TT} \tfrac{1}{C(\TT)}\,  e^{\k_0 N_0} A(\TT)\, ,
\ee
which is related to the expectation value as a function of $\k_3$ via
\be
\la A \ra = \frac{1}{Z_{\rm gc}} \sum_{N_3} e^{-\k_3 N_3} Z_{\rm c}(N_3)\, \la A \ra_{N_3} \, .
\ee
%

\subsection{The simulations}

Simulations were performed using the Markov-chain Monte-Carlo technique.  An adaptation of some previously existing code for the Monte Carlo simulations (used in \cite{Ambjorn:2000dja}) was used for this purpose.  The code generates a finite set of sample configurations $\{\TT_1,...,\TT_M\}$ according to the probability distribution $\PPP(\TT) = \f{1}{Z} e^{-S(\TT)}$.  We approximate the expectation value of an observable by its arithmetic mean across these samples:
\be
\la \OO \ra \approx \f{1}{M} \sum_{j=1}^M \OO(\TT_j).
\ee

As usual in 2+1D CDT simulations, the spacetime topology was fixed to $S^2 \times S^1$, \textit{i.e.} spherical spatial sections and cyclical time.\footnote{Simulations with different boundary conditions in the time direction have been performed by Cooperman and Miller in \cite{Cooperman:2013mma} and their results give us confidence that the results presented here are not affected by our choice.} Values of $N_3$ up to a maximum of 200k (meaning $2 \times 10^5$) were studied, although some errors for the larger values of $N_3$ are greater since less configurations could be generated to be averaged over, within practical time constraints.  All simulations were carried out with coupling constant $\kappa_0=5$, in the phase analogous to phase $C$ in 3+1D, where previous evidence points to the emergence of well-behaved geometry.  The total number of time-steps was set to $T=96$.

\subsection{The volume data}

The observable we study here is the volume of the spatial slices, which in the (2+1)-dimensional CDT model corresponds to the number of spatial triangles $N_2(i)$ as a function of discrete time $i$. 
Because the triangulation is connected we always have $N_2(i)>0$. Furthermore, because we restrict to simplicial manifolds, the smallest triangulation of a two sphere has four triangles, giving
\be \label{minN2}
N_2(i) \geq n_\eps =4\, .
\ee

\begin{figure}[ht]
\centering 
\includegraphics[width=13cm]{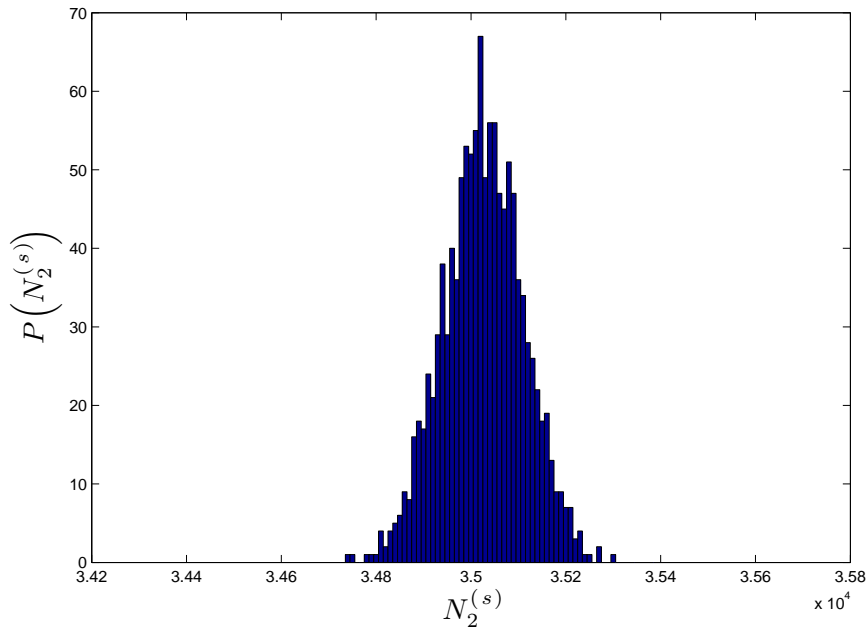} 
\caption{\small{The distribution of the total number of spatial triangles $N_2^{(s)}$ (grouped in bins of size 10), for $N_3$=100k. The expectation value is $\la N_2^{(s)}\ra =35026$, with standard deviation $\s=85$.}}
\label{Fig:N_2-tot}
\end{figure}

Another possible observable is the volume of the spatial slices at half-integer values of time, which amounts to a weighted sum of the number of (3,1) and  of (2,2) tetrahedra between slices $i$ and $i+1$. We expect that in the phase of extended geometry (where $N_{(3,1)}\sim 2 N_{(2,2)}$) any such differences in definitions of volume as a function of time should be irrelevant in the continuum limit.
Note that a triangle is always shared by two tetrahedra, so that, for the number $N_2^{(s)}$ of spatial triangles, we have
\be \label{N2s}
N_2^{(s)} = \sum_{i=1}^T N_2(i) = \f12 N_{(3,1)}\, ,
\ee
which in the extended phase we expect to be roughly one third of the total volume $N_3$.
For the value of $\k_0$ we used,  the distribution of $N_2^{(s)}$ is very peaked (for $N_3=$100k the relative standard deviation is only $0.2\%$, see Fig.~\ref{Fig:N_2-tot}), and we find that $N_{(2,2)}$ is just slightly smaller than a third of the total volume.

\begin{figure}[ht]
\centering 
\includegraphics[width=13cm]{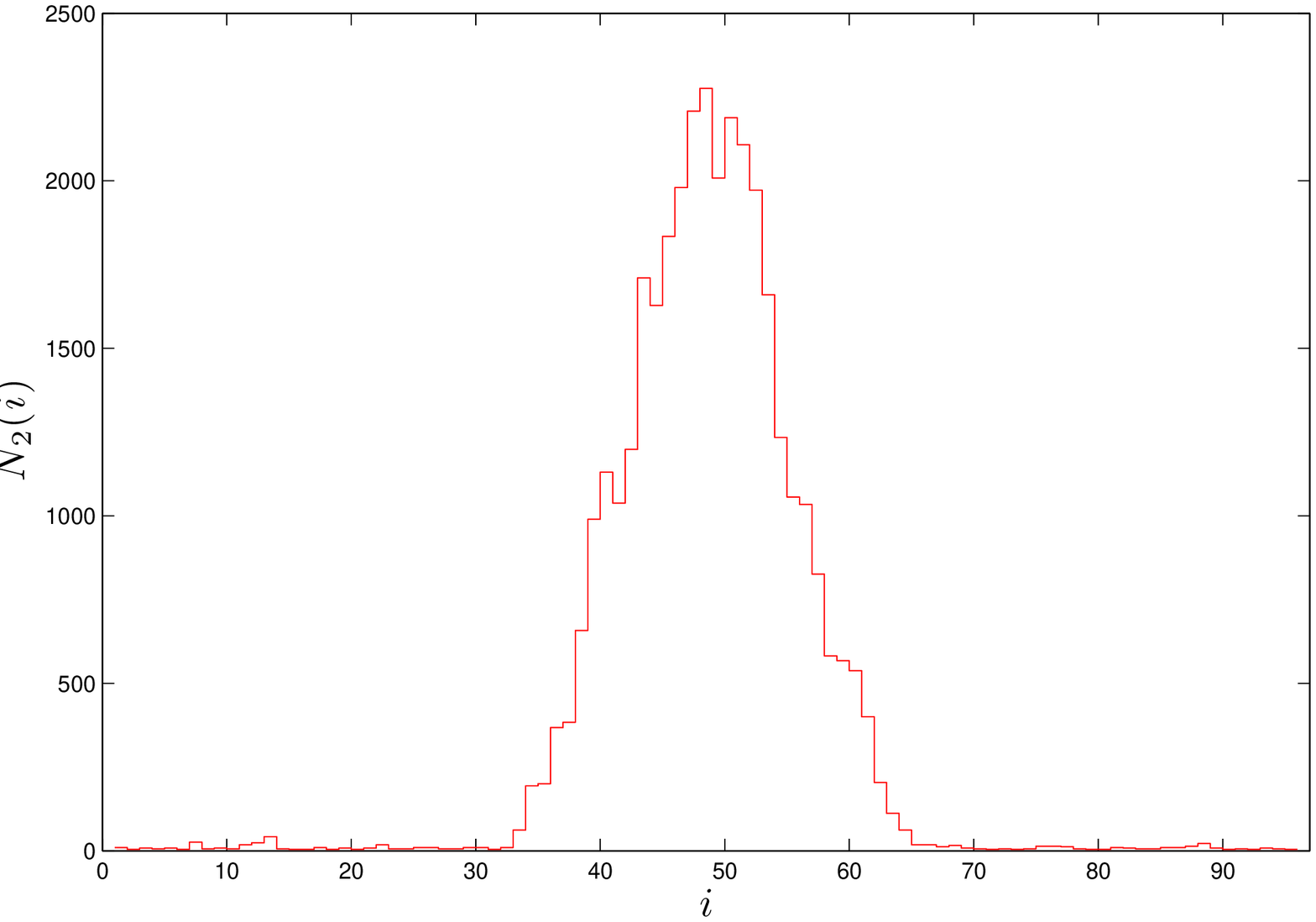} 
\caption{\small{The spatial area $N_2(i)$ as a function of time, from a single configuration taken at random in the data set of MC simulations for $N_3$=100k. 
}}
\label{Fig:hist-N_2}
\end{figure}

\begin{figure}[ht]
\centering 
\includegraphics[width=13cm]{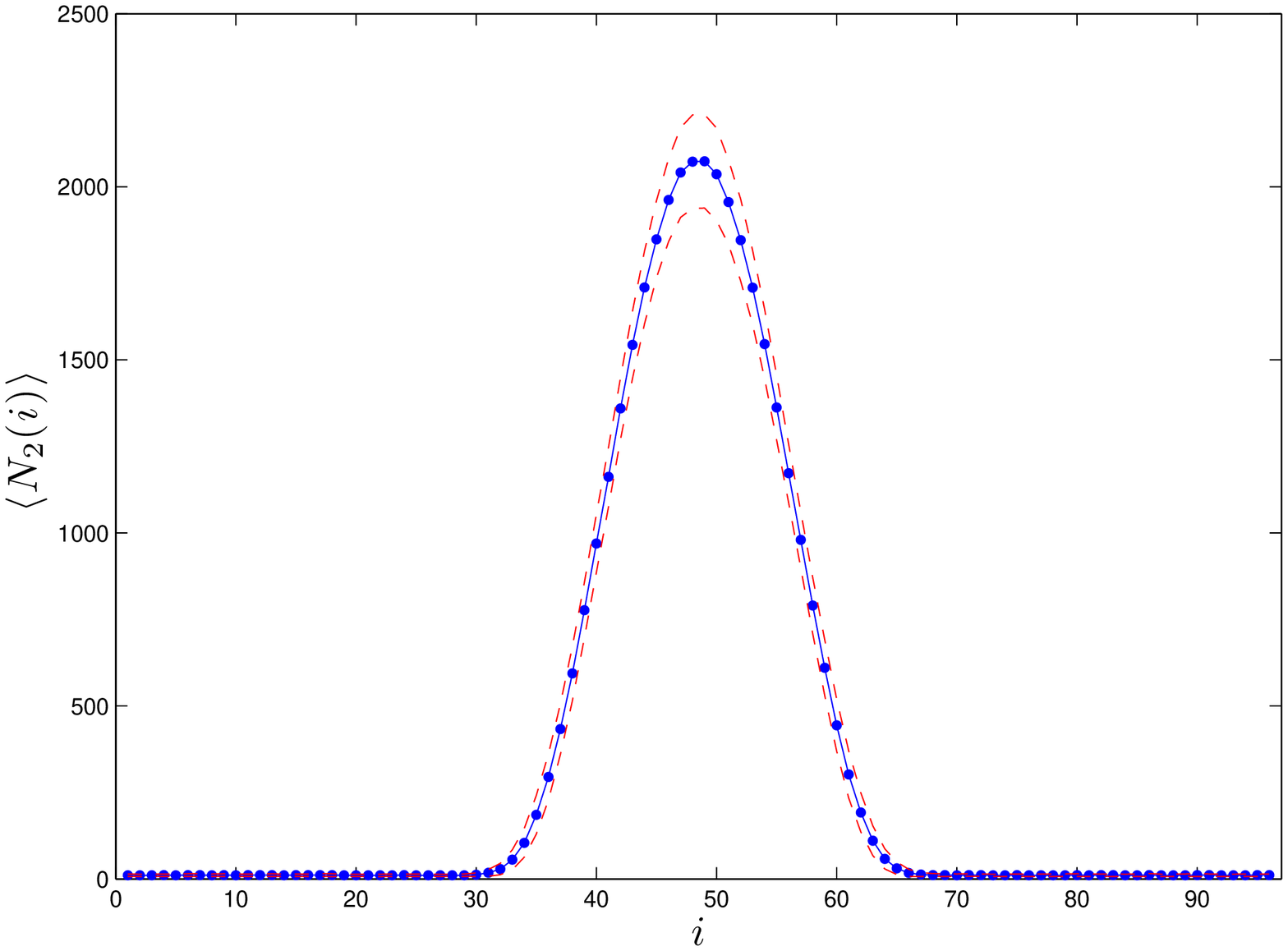} 
\caption{\small{The mean spatial area as a function of discrete time, for $N_3$=100k. The blue dots represent the mean area $\la N_2(i)\ra$  of the volume data at time $i$, while the blue line is just an interpolation curve. Error bars (estimated for example by the jackknife method) are roughly of the size of the dot radii, or smaller, and thus we chose to omit them for clarity (this is done throughout the paper).
The dashed red lines are interpolating curves for $\la N_2(i)\ra \pm \s/2$, where $\s=\sqrt{\la (N_2(i))^2\ra-\la N_2(i)\ra^2}$ is the size of the fluctuations.}}
\label{Fig:N_2}
\end{figure}

In  Fig.~\ref{Fig:hist-N_2}  we show the volume profile $N_2(i)$ from a snapshot of a MC simulation.
One notices immediately  a phenomenon of spontaneous (translational) symmetry breaking: the MC configuration shows a condensation of the volume around a specific time. Averaging over MC configurations, the translational symmetry gets restored, but in this way we loose information about the typical configuration that dominates in the partition function.
Therefore, before being able to do any meaningful analysis we have to find the center of volume $t_{CV}(j)$ for each MC configuration $j$, and we have to shift time so that $t'_{CV}(j)=T/2$ for every configuration in the new time variable. We performed this operation following the method given in \cite{Ambjorn:2008wc,Gorlich:2011ga}. 
Once the data are centered in this way, it makes sense to study the average of  $N_2(i)$. A plot of $\la N_2(i)\ra$,  together with fluctuations, is displayed in Fig.~\ref{Fig:N_2}.


The latter is the observable which is central to this work. The volume profile has a characteristic extended part (typically referred to as the \textit{blob} or \textit{droplet}), and a long flat tail (referred to as the \textit{stalk}).  Within the latter, the spatial volume is very close to its kinematical minimum, $\f1m \sum_{i=1\ldots m}^{i\in {\rm stalk}}\la N_2(i)\ra \equiv n_s \sim 10$, and it is independent of the total volume. Most of the total volume is therefore concentrated in the blob.
We will discuss in the following sections how to explain this condensation phenomenon, and which function best describes the volume profile.

\subsection{The continuum limit}
\label{s:c_limit}

We conclude this section by discussing how the continuum limit $a\to 0$ is investigated on the basis of the simulation data.
All observables and couplings in the simulations are given as dimensionless numbers. Length dimensions are introduced by multiplying the quantity of interest by the appropriate power of the cutoff $a$.
For example, we can write $\t = a \, \a\,   T$ for the time interval, where we have introduced a parameter $\a>0$ to scale the timelike edges with respect to the spacelike ones. Next, we can write $V_2 = \f{\sqrt{3}}{4} a^2 N_2$ for the volume of a slice (the numerical prefactor being the area of an equilateral triangle of unit side),  $V_3 = v_{(3,1)}(\a) a^3 N_{(3,1)} + v_{(2,2)}(\a) a^3 N_{(2,2)}$ for the total volume, etc. Here, $v_{(3,1)}(\a)$ and $v_{(2,2)}(\a)$ stand for the volume of the $(3,1)$ and $(2,2)$ simplices with spatial edges of length one, and time edges of length $\a$ (they both coincide with the equilateral tetrahedron for $\a=1$, with volume $v_3=1/6\sqrt{2}$, see  \cite{Ambjorn:2012jv}).

In particular, by keeping $V_3 \sim a^3 N_3$ fixed, we have the fundamental scaling relation $a\sim N_3^{1/3}$.
Therefore we construct the continuum limit by rescaling all quantities by the appropriate power of $N_3^{1/3}$, according to their dimension, and taking a larger and larger volume.
Without any fine tuning of the dimensionless couplings, all the couplings having positive (negative) length dimension will go to zero (infinity) as $a\to 0$. This is for example the case for Newton's constant, which has the dimension of length in 2+1D and thus is expected to go to zero in the naive continuum limit (in four spacetime dimensions, where Newton's constant has dimension of length to the second power, this has been observed in \cite{Ambjorn:2008wc}). 
A finite coupling in the continuum could hopefully be attained by fine tuning the bare coupling to a second order phase transition.\footnote{This is the general picture relating continuum quantum field theory (with propagating degrees of freedom) and lattice field theory, via the fine tuning of the latter to a second-order phase transition lying in the universality class of some fixed point of the renormalisation group. In the case of gravity this would lead to either the realisation of an asymptotic safety scenario \cite{Weinberg:1980gg,Niedermaier:2006wt,Percacci:2007sz,Litim:2011cp,Reuter:2012id}, or the existence of a Lifshitz critical point  \cite{Horava:2008ih,Horava:2009uw} (or perhaps some novel scenario made possible by non-standard features of the gravitational path integral). The difference is in general between trivial and nontrivial fixed points (although an interacting Lifshitz critical point is also possible), and it could in principle be discerned by studying the critical properties associated to a phase transition. In this respect, it is very encouraging that a second-order phase transition has been identified in CDT \cite{Ambjorn:2011cg,Ambjorn:2012ij}.}
However, some dynamically generated quantities, for example the width or amplitude of the droplet, might show scaling with $N_3$, naturally leading to finite dimensionful quantities in the limit in which the cutoff is removed.
\begin{figure}[ht]
\centering 
\includegraphics[width=13cm]{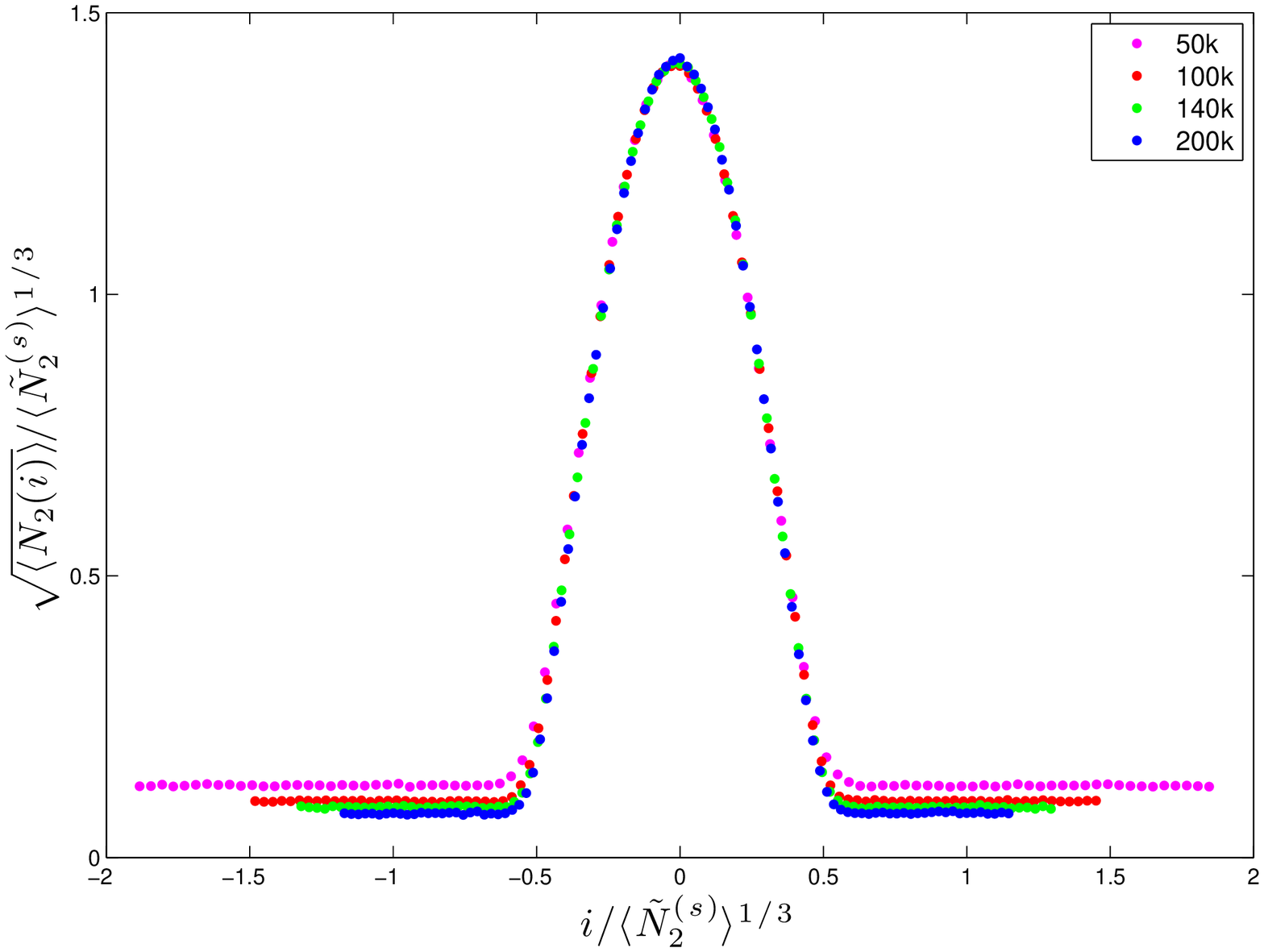} 
\caption{\small{The square root of the mean spatial area as a function of discrete time (shifted so that the peak is at the origin). Both the mean area and the time variable are rescaled with the appropriate power of the average total volume $\tilde N_2^{(s)} = N_2^{(s)} - n_s T$, in order to display scaling.}}
\label{Fig:scaling}
\end{figure}
This is precisely what Fig.~\ref{Fig:scaling} shows. Here we have plotted  $\la N_2(i)\ra^{1/2}/ \langle \tilde N_2^{(s)}\rangle^{1/3}$ as a function of $i/ \langle \tilde N_2^{(s)}\rangle^{1/3}$, for different data sets corresponding to different total volumes. 
Following an analogue of the procedure used in \cite{Ambjorn:2005qt}, we used   
\be \label{red-vol}
\la \tilde N_2^{(s)}\ra = \la N_2^{(s)} \ra - n_s T\, ,
\ee
instead of $\la N_2^{(s)}\ra$ (or $N_3$, which as we saw, is proportional to it) in the rescaling because we know that the volume in the stalk does not scale.
The plot clearly shows that the superposition is extremely good inside the droplet, while in the stalk the rescaled volume goes to zero for growing $N_3$.
Another dynamical quantity that goes to zero in the continuum is the size of the fluctuations around the average, which we expect  to be controlled by Newton's constant.

\section{Balls-in-boxes models}
\label{Sec:BIB}

The statistical models known as  balls-in-boxes (or zero-range process in the nonequilibrium version) are an interesting and versatile class of models which have been extensively studied in the statistical mechanics literature (see for example \cite{Evans-review}).
They have also been used as mean field models of (non-causal) dynamical triangulations \cite{Bialas:1996eh,Bialas:1997qs,Bialas:1998ci}.
More recently, they have been studied in  \cite{Bogacz:2012sa} as effective models for the spatial volume dynamics of CDT in 3+1 dimensions. We briefly review in this section some of their relevant properties, and their connection to CDT.

A  balls-in-boxes (BIB) model is defined as a one-dimensional lattice with $T$ sites (boxes) to each of which is associated an integer number $m_i\geq m_{\rm min}>0$ (the number of balls in box $i\in\{1,2, ..., T\}$). The total number of balls is fixed to be $M$.
The canonical partition function of the statistical model is given as
\be\label{Z_BIB}
\begin{split}
Z_{\rm BIB}(T,M) &= \sum_{m_1=m_{\rm min}}^M ...  \sum_{m_T=m_{\rm min}}^M \d_{M,\sum_i m_i } \prod_{i=1}^T g(m_i,m_{i+1}) \\
 &= \sum_{\{m_j\} } e^{-S[\{m_j\}]}  \d_{M,\sum_i m_i } \, ,
 \end{split}
\ee
with $m_{T+1}=m_1$. The last expression highlights the interpretation of such models as discretised one-dimensional path integrals, subjected to the constraint
\be \label{constraintM}
\sum_{i=1}^T m_i  = M \, .
\ee
The weight function $g(m,n)$ (or the action $S[\{m_i\}]$) defines the particular model. In the standard BIB models there is no nearest-neighbour interaction, meaning that $g(m_i,m_{i+1}) =g(m_i)$; the model above is a generalisation studied in \cite{Evans-prl,Waclaw:2009zz}.
For the effective description of the spatial volume dynamics in CDT, we will see that the action depends on the dimension of the CDT model.

Often, in particular for an analytical approach, it is useful to work in the grand canonical ensemble, for which the partition function reads
\be \label{Z_BIB-gc}
\begin{split}
Z_{\rm BIB-gc}(T,z) &= \sum_M Z_{\rm BIB}(T,M)\, z^M \\
   &= \Tr \, \hat T^T \, ,
\end{split}
\ee
where we introduced the transfer matrix
\be
\hat T_{m,n} = z^{(m+n)/2} g(m,n) \, .
\ee
In the light of this relation, $g(m,n)$ is sometimes referred to as {\it reduced transfer matrix}.

The interesting feature of these models, at least in this context, is that they can exhibit a condensation phenomenon. In the original BIB models, with $g(m_i,m_{i+1}) =g(m_i)$, this means that for certain values of the parameters the model enters into a phase dominated by configurations completely localised at one (random) site. 
The mechanism behind such condensation remains similar in the more general models, but the nearest-neighbour interaction allows the condensate to spread over a region whose width scales with a power of the total volume $M$.
It is this type of condensation which provides the basis for an explanation of the droplet configuration in CDT based on the much simpler BIB models.

In the remaining part of this section, we review the relation between BIB and CDT models in 1+1 and 3+1 dimensions, and we discuss the problems of the $(2+1)$-dimensional case. The purpose of such review and discussion is to set the stage for the ensuing analysis of Sec.~\ref{Sec:miniHL}, and introduce some ideas and results that will be used there.

\subsection{(1+1)-dimensional CDT}

The $(1+1)$-dimensional model of CDT \cite{Ambjorn:1998xu} is precisely of the form \eqref{Z_BIB}, with $m_i=l_i$ giving the length of the spatial slice. In this case,
\be
g(l_i,l_{i+1}) =  \f{ (l_i + l_{i+1} )! }{ l_i! \, l_{i+1}! } \, ,
\ee
counts exactly the number of triangulations of a strip with boundary lengths $l_i$ and $l_{i+1}$, with open boundary conditions.\footnote{It basically counts the number of ways we can place $l_{i+1}$ balls in $l_i+1$ boxes. Therefore the $(1+1)$-dimensional model of CDT is a BIB model whose reduced transfer matrix is defined by an auxiliary BIB model.}

Using Stirling's formula one finds that, for large $l_i$ and $l_{i+1}$, and small  $(l_{i+1}-l_i)/(l_i + l_{i+1})$,
\be \label{2d-transfer}
g(l_i,l_{i+1}) \sim 2^{l_i + l_{i+1} } e^{-\f{(l_{i+1}-l_i)^2}{l_i + l_{i+1}}} \, ,
\ee
where in the exponent we recognise a discrete version of the typical kinetic term that arises in minisuperspace models of general relativity (or as we will see, Ho\v rava-Lifshitz gravity).\footnote{The role of higher order terms in the Stirling approximation has been studied in a recent work \cite{Ambjorn:2015gea}, where it was shown that including enough matter fields the leading order correction (a logarithmic potential term in the exponent of \eqref{2d-transfer}) becomes important and leads to a phase transition to a droplet phase, similar to the one we will describe below.}

The model is exactly solvable \cite{Ambjorn:1998xu,DiFrancesco:1999em}, and it has recently been linked to (1+1)-dimensional HL gravity \cite{Ambjorn:2013joa}.
The latter reduces to a one-dimensional action for the length $L(t)$ of the slices \cite{Ambjorn:2013joa}, which in proper-time gauge reads
\be \label{S_2d}
S_{\rm (1+1)-HL} = \int_{-\f{\t}{2}}^{\f{\t}{2}} dt\,
\f{\dot L^2(t)}{4 L(t)}
 \, ,
\ee
and whose Lagrangian can be interpreted as the continuum limit of the exponent in \eqref{2d-transfer}.

It is well known (see for example \cite{Mattei:2005cm} and references therein) that in a path integral quantisation of gravity in the proper-time gauge we loose the Hamiltonian constraint, unless we integrate over the total proper time. Such integration is not performed in CDT when computing finite time propagators (by the definition of such observables), or when doing simulations with periodic boundary conditions in time,  for obvious practical reasons.
But by no means should one conclude from this alone that CDT is not actually recovering GR: in principle, nothing forbids us from doing the integral over time at a later stage.  Indeed, in 1+1 dimensions this integral can be carried out exactly, leading to a solution of the Wheeler-DeWitt equation \cite{Ambjorn:1998xu}.
What we want to stress here is that in order to make contact between continuum models and CDT results with fixed total time, one should not try to impose the Hamiltonian constraint in the former.
Bearing this in mind, we can try to see what a semiclassical analysis of \eqref{S_2d} tells us. Solving the equations of motion under the constraint $V_2=\int_{-\f{\t}{2}}^{\f{\t}{2}} dt L(t)$, and with periodic boundary conditions in time, one finds either constant or oscillating solutions.
The constant solution is in fact unique, due to the volume constraint: $L(t)=V_2/\t$. The oscillating solutions form a discrete set, due to the periodicity condition (we can have one or multiple oscillations, but always an integer number of them). 
Plugging these solutions into the action \eqref{S_2d} it is easy to check that the constant solution has the least action (it evaluates to zero) and therefore it must dominate the path integral. 
Note that the Hamiltonian constraint would fix the amplitude independently of the total time $\t$, and therefore we would have periodic solutions only for special values of $\t$.
We could consider also droplet configurations of the kind that we will introduce in higher dimensions, but the constant solution would still dominate over them, and since such analysis is a particular case of the one we perform in Sec.~\ref{Sec:miniHL} we omit it here.

The dominance of the constant solution is in complete agreement with the MC snapshots from numerical simulations \cite{Ambjorn:1999gi}, which, unlike the higher-dimensional models, show no sign of translational symmetry breaking.\footnote{The analytical solution \cite{Ambjorn:1998xu} gives $\la L(t) \ra \propto 1/\sqrt{\L}$ ($\L$ being the cosmological constant in the grand canonical ensemble), but of course in the case of periodic boundary conditions a simple average will always lead to a translational-invariant result, and that's why we refer to MC snapshots rather than averages. One could break explicitly the translational symmetry by introducing initial and final boundaries of fixed length, and check that for large $\t$ the bulk is approximately constant, but this would be a long and unnecessary parentheses here.
More importantly, one should remember that in 1+1 dimensions $\sqrt{\la L^2 \ra-\la L \ra^2} \propto 1/\sqrt{\L}$, i.e. fluctuations have the same magnitude as the average configuration, thus hiding any possible classical behaviour of the latter.}

\subsection{The effective model for (3+1)-dimensional CDT}

In  \cite{Bogacz:2012sa}, Bogacz et al. studied a BIB model, which is basically a discretised version of the minisuperspace model corresponding to (3+1)-dimensional general relativity.
It was found that this very simple model can account for many of the observed features of CDT in four dimensions, including its rich phase diagram. In particular, a droplet phase was found, which has remarkable similarities to the extended phase of CDT (of course, the comparison is limited to the behaviour of the spatial volume against time).

The model is defined by the following reduced transfer matrix,
\be \label{BIB4d}
g(m_i,m_{i+1}) = \exp\left( -c_1 \f{2 (m_{i+1}-m_i)^2}{m_i + m_{i+1}} - c_2 \f{m_i^{1/3} + m_{i+1}^{1/3}}{2} \right)\, .
\ee
In the continuum this corresponds to the following action for the volume  $V_3(t)$ of the 3-dimensional spatial slices,
\be \label{S_mini4d}
S_{\rm (3+1)-mini} = \f{1}{2 G} \int_{-\f{\t}{2}}^{\f{\t}{2}} dt \, \left( c_1 \f{\dot V_3^2(t)}{ V_3(t)} + c_2 V_3^{1/3}(t) \right) \, ,
\ee
which is precisely of the type obtained from a minisuperspace reduction of general relativity (where $c_1=1/N$, and $c_2=9 (2\pi^2)^{2/3} N$, $N$ being the lapse function).
This is the action that was conjectured from the very beginning by Ambj\o rn \textit{et al.}~as an effective description for the extended part of the universe (or blob) in their simulations \cite{Ambjorn:2004pw,Ambjorn:2005qt}, a conjecture which was further corroborated over the years \cite{Ambjorn:2007jv,Ambjorn:2008wc,Ambjorn:2011ph}.
The novelty in \cite{Bogacz:2012sa} was the suggestion that the same effective action can explain much more than just the dynamics inside the blob.

One important difference between the usual minisuperspace model of general relativity and the BIB model, is that in the latter there is no analogue of the lapse to be integrated in the partition function and neither there is a summation/integration over $T$. As a consequence, there is no Hamiltonian constraint to be imposed in the semiclassical analysis. 
Above, we emphasised that the same situation should be expected in CDT, where the distance between one spatial slice and the next is constant (i.e. the lapse is constant) and the total time extension of the universe is fixed in all simulations to date.
This is also supported by the strong evidence from numerical simulations that the BIB model is a good effective description for CDT.

The equations of motion derived by varying \eqref{S_mini4d} with respect to $V_3(t)$ are
\be \label{eom4d}
c_1\left( \left( \f{\dot V_3}{V_3} \right)^2 -2 \f{\ddot V_3}{V_3}  \right)  + \f{c_2}{3} \f{1}{V_3^{2/3}} - \L =0 \, ,
\ee
where the cosmological constant $\L$ is introduced as a Lagrange multiplier, to be fixed by imposing the volume constraint. If we were to impose also the Hamiltonian constraint $H\equiv c_1 \dot V_3^2(t)/V_3(t) - c_2 V_3^{1/3}(t) +\L V_3(t) = 0$, the combined system of equations would reduce to a first order differential equation (by deriving the Hamiltonian constraint with respect to time, and eliminating $\ddot V_3(t)$ between the two equations).
As such, its solutions would have only one free integration constant, which could be fixed for example by demanding that the maximum of $V_3(t)$ be at $t=0$.
The solution would then be the ``$\cos^3$'' solution discussed by Ambj\o rn et al.  in \cite{Ambjorn:2007jv,Ambjorn:2008wc}.
However, without the Hamiltonian constraint the equation remains second-order, and thus there is one more free parameter. 

Bogacz et al. fix the free parameter by minimisation of the on-shell action, as we have done above for the (1+1)-dimensional case.
For such a minimization one does not need to restrict to class $C^2(S^1)$ functions, since for a well defined action \eqref{S_mini4d} it is sufficient that $V_3(t)\in C^1(S^1)$. This allows the authors of \cite{Bogacz:2012sa} to consider droplet configurations that are expected on the basis of simulations and heuristic arguments.\footnote{One should bear in mind that this sort of analysis is not aimed at reproducing the detailed profile of CDT at the junction between blob and stalk. In the junction region of the droplet we expect the effect of subleading terms in the action to be non-negligible. Furthermore, the time-interval mesh in the CDT simulations is not fine enough to reveal much about the smoothness of the average profile at such junction.}
Using $V_4 = \int_{-\f{\t}{2}}^{\f{\t}{2}} dt V_3(t)$ for the total volume in the continuum, they obtain the following expression for the dominant contribution to the path integral in a particular region of the the phase diagram:\footnote{A detailed analysis of the parametric conditions under which such configuration dominates, goes beyond the scope of the short review we are making here, and it is very similar to the one we will make in detail for the $(2+1)$-dimensional case in Sec.~\ref{Sec:miniHL}.}
\be \label{fullsol4d}
\bar V_3(t) = \begin{cases} \f{3 \om V_4}{4} \cos^3(\om t) \, ,  &  \text{for } t \in [-\f{\pi}{2\om},+\f{\pi}{2\om}]\, ,  \\  0  \, ,  &  \text{for } t \in [-\f{\t}{2},-\f{\pi}{2\om})\cup (+\f{\pi}{2\om},+\f{\t}{2}]\, ,  \end{cases}
\ee
where 
\be \label{frequency4d}
\om = \f{\sqrt{2} }{3 V_4^{1/4} } \left( \f{c_2}{c_1} \right)^{3/8}\, .
\ee
Notice that $V_3(t)=0$ obviously minimises the action \eqref{S_mini4d} for positive $c_1$ and $c_2$. However, alone it would fail to satisfy the volume constraint, and therefore a balance between the zero and the  ``$\cos^3$'' solutions wins the energy balance, resulting in a condensation. Interestingly, \eqref{fullsol4d} and \eqref{frequency4d} correspond to the solution obtained by imposing also the Hamiltonian constraint, but in our opinion this is a mere coincidence. 

The crucial point to be made here is that the presence of a potential term in \eqref{S_mini4d} allows non-constant configurations such as \eqref{fullsol4d} to dominate the path integral.  Thus, the conclusions derived in this case are qualitatively different from those derived from \eqref{S_2d} in the (1+1)D case.

The result is very interesting because it shows how the reduced model in 3+1 dimensions not only reproduces the extended part of the universe, but also its stalk, and it gives a prediction for their relative time extension.


\subsection{The GR-inspired effective model for (2+1)-dimensional CDT}
\label{Sec:3d_troubles}

In 2+1 dimensions, the classical minisuperspace action derived from GR in the proper-time gauge is exactly of the same form as \eqref{S_2d}, but with the length $L(t)$ replaced by the volume (or area) of the spatial slices $V_2(t)$. As in that case, in the presence of a volume constraint  $V_3 = \int dt V_2(t)$ this gives as a solution of the equations of motion a classical volume profile  $V_2(t)\sim \cos^2 (\om t)$ which is compatible with the one observed in CDT, and as such it has been suggested as an effective action for CDT \cite{Ambjorn:2000dja,Ambjorn:2002nu} (see also \cite{Cooperman:2013mma} for a more detailed discussion in this (2+1)-dimensional context).

In the light of the recent results on BIB models for CDT, the similarity of the minisuperspace GR action to  \eqref{S_2d} immediately raises the question of how the BIB models could ever explain the important differences between (1+1) and (2+1)-dimensional CDT. If the (1+1)-dimensional case is exactly a BIB model, and the (3+1)-dimensional case is well described (in its spatial volume dynamics) by a BIB model, should we not also expect a good approximation for the (2+1)-dimensional model? 

Of course an important difference between 1+1 and 2+1 dimensions is in the scaling of dimensionful quantities. Most importantly, Newton's constant is dimensionless in 1+1 dimensions, while it has dimension of length in 2+1D. As a consequence, in the (naive, not fine-tuned) continuum limit, Newton's constant (and with it the fluctuations around the average volume of the slices) scales to zero in the latter case, while it stays constant in the former. In 1+1 dimensions, as there are no other scales besides the cosmological one, the size of the fluctuations in the continuum limit is as large as the expectation value, thus blurring any classical behaviour. On the contrary, in 2+1 dimensions the fluctuations go to zero and the classical (mean field) behaviour should dominate.

However, regardless of how the fluctuations behave, it turns out that the GR-inspired action fails in reproducing the CDT results in an important way.
Repeating the analysis of Bogacz et al. for 2+1 dimensions we simply have to set $c_2=0$ in the previous subsection (or just recall what we have said about the (1+1)d case).
From \eqref{frequency4d} we then find that $\om= 0$, and the width of the droplet diverges.
Being more careful,  \eqref{frequency4d} does not hold in this case, as it would violate the condition $\pi/\om<\t$ which is to be assumed in \eqref{fullsol4d}.
But we have already explained what happens in the (1+1)-dimensional case. Going to 2+1 dimensions we simply have to replace  $V_2\to V_3$ and $L(t)\to V_2(t)$.
The droplet solution which minimises the action is obtained for $\om=\pi/\t$. However, it is easily checked that in such a case the action is strictly positive, while for $V_2(t)=V_3/\t$ the action vanishes. We conclude that a BIB model inspired by (2+1)-dimensional GR would predict a constant average profile for the two-dimensional volumes.
This is also supported by the numerical simulation of   \cite{Bogacz:2012sa}, as for $c_2=0$ and $c_1>0$ the model defined by \eqref{BIB4d} lies in the correlated fluid phase, not the droplet phase.

We are left with the challenge of explaining the droplet condensation of 2+1D CDT as a BIB-type condensation.  This also provides us with an extraordinary opportunity to test corrections to the GR effective action. In higher dimensions, such corrections are expected to be subdominant with respect to the the linear spatial curvature term coming from GR (the one multiplied by $c_2$ in \eqref{S_mini4d}). 
Fortunately, however, the 2+1D case is an exception to this, because this curvature term is topological (it just gives the Euler character of the spatial manifold), and hence drops out of the story, making higher order corrections relevant.

\section{A minisuperspace model of Ho\v{r}ava-Lifshitz gravity in 2+1 dimensions}
\label{Sec:miniHL}

We seek here a continuous effective description of the CDT model, at least for the dynamics of the spatial volume, in the spirit discussed in the previous section. To that end we could start directly with an ansatz for a minisuperspace action.  However, this would leave us with no guidance and too much freedom. Instead, we start by postulating a full (i.e. not minisuperspace) action, based on general principles and expectations.

Motivated by the evidence accumulated in recent years \cite{Horava:2009if,Benedetti:2009ge,Ambjorn:2010hu,Budd:2011zm,Ambjorn:2013joa}, we conjecture that the full CDT model falls into a Ho\v{r}ava-Lifshitz (HL) type of universality class \cite{Horava:2008ih,Horava:2009uw}, i.e. that a preferred foliation survives in the continuum limit\footnote{At least in a naive continuum limit. We cannot exclude \textit{a priori} that with some \textit{ad hoc} (and nontrivial) fine tuning of the bare couplings, a continuum limit with enlarged symmetry might be achieved (on the contrary, we believe this to be a reasonable possibility).
On the other hand, a modification of CDT has been introduced in \cite{Jordan:2013iaa}, where the foliation structure at the discrete level is relaxed to some extent. The large scale results appear to be very similar to the standard CDT results, and we take this as evidence of universality, with a foliation emerging in the continuum limit.}.
In other words we begin by constructing an action in terms of geometric invariants respecting all the symmetries compatible with a preferred foliation. These include in particular spatial diffeomorphisms, which we expect not to be broken in CDT.
  In HL gravity time reparametrisation is also a postulated symmetry, whose implementation requires the introduction of a lapse function, and therefore it leads to a Hamiltonian constraint. As we discussed previously, in our opinion there is currently no reason why such a constraint should be required in interpreting the continuum limit of CDT data: the discrete model clearly has a constant lapse, and since no integration is performed over the proper time, no Hamiltonian constraint is enforced. Of course nothing would forbid one to do such an integration at a later stage, but as here we want to make contact with the available data, in what follows we will set $N$ to a constant, and we will not impose any Hamiltonian constraint.\footnote{Note that in any case HL gravity generally contains an additional scalar degree of freedom as compared to general relativity, so the non-imposition of the Hamiltonian constraint should not alter the counting of degrees of freedom.}

The action that we consider as an ansatz in this section is a particular case of the projectable version of Ho\v rava-Lifshitz gravity. The latter is characterised by a spatially constant lapse function, $N=N(t)$, and its most generic $z=2$ action in $2+1$ dimensions reads
\be\label{actionHL}
S_{\rm (2+1)-HL} =  \frac{1}{16\pi G} \, \int d t\, d^2 x N \sqrt{g} \, \left\{ \s( \l\, K^2 - K_{ij}K^{ij} )- 2\, \L + b\, R -\g\, R^2  \right\}\, ,
\ee
where $g$ is the determinant of the spatial metric, $R$ its Ricci scalar, $K_{ij}$ the extrinsic curvature of the leaves of the foliation, and $K$ its trace. The parameter $\s=\pm 1$ is introduced for the sake of generality. $G$ is Newton's constant, and $\L$ is the cosmological constant, while $\l$, $b$ and $\g$ characterise the deviation from full diffeomorphism invariance (for $\l=b=1$ and $\g=0$ the Lagrangian reduces to $\cR-2\L$, where $\cR$ is the Ricci tensor of the whole spacetime in Euclidean ($\s=1$) or Lorentzian ($\s=-1$) signature).
In HL gravity the exponent $z$ refers to the number of spatial derivatives appearing in the inverse propagator of the free theory, and it is known that $z\geq d$ in $d+1$ dimensions is needed for renormalisability. 
In CDT however a renormalisation group analysis has only just begun \cite{Ambjorn:2014gsa,Cooperman:2014owa}, and we are currently not in a position to say anything about the presence of a $z=2$ Lifshitz point in (2+1)-dimensional CDT. Therefore, our point of view on \eqref{actionHL} is that of effective field theory: given the symmetries we have assumed,  \eqref{actionHL} contains the leading terms in an expansion of the effective potential in operators of increasing dimension.

Next, we consider a mini-superspace reduction of  \eqref{actionHL}  for constant $N$, in which we restrict also to vanishing shift vector (otherwise implicitly contained in the extrinsic curvature tensor), and where for the spatial metric we take
\be \label{metric}
g_{ij} = \phi^2\, \hat g_{ij}\, ,
\ee
$\hat g_{ij}$ being the standard metric on the unit sphere.
The function $\phi=\phi(t)$ is a time-dependent scale factor, determining the area of a spatial slice:
\be \label{V2}
V_2(t)=\int d^2 x \sqrt{g} = 4\pi \phi^2(t)\, .
\ee

Substituting \eqref{metric} in \eqref{actionHL}, we find the mini-superspace action
\be \label{Smini}
\begin{split}
\bar S_{\rm (2+1)-mini} & = \f{ N}{2G} \int_{-\f{\t}{2}}^{\f{\t}{2}} dt \, \left\{   \f{\s (2\l-1)}{N^2}  \dot{\phi}^2 -  \L \phi^2 + b - \f{2\g}{\phi^2}  \right\} \\
& =  \f{ N}{2G} \int_{-\f{\t}{2}}^{\f{\t}{2}} dt \, \left\{  \f{ \s (2\l-1)}{16\pi N^2}   \f{\dot{V_2}^2}{V_2}  - \f{\L}{4\pi} V_2 +b -\f{8\pi\g}{V_2}  \right\} \, ,
\end{split}
\ee
where we assume periodic boundary conditions with period $\t$. 
The kinetic term is positive definite for $\s (2\l-1)>0$, which we will assume from this point onwards.
On the other hand, because we want oscillating (i.e. periodic) real solutions, we are forced to take the cosmological and the $R^2$ terms with an opposite sign with respect to the kinetic term ($\L>0$ and $\g>0$), effectively leading to an action with Lorentzian signature, $\bar S_{\rm (2+1)-mini}=\int dt [\cL_{\rm kin}-\cL_{\rm pot}]$ (but if gravitons were present, and if the $R$ term was not topological, we would not be able to interpret it in such a way).
As a consequence the action would seem to be unbounded from below. 
However, the point of view we adopt here, in the spirit of the CDT simulations, is that of fixing the total volume of spacetime, and thus viewing the cosmological constant as a Lagrange multiplier. 
That is, we impose the constraint
\be \label{constr}
4\pi N \int_{-\f{\t}{2}}^{\f{\t}{2}} dt\, \phi^2(t) = V_3 \, ,
\ee
by adding to the action \eqref{Smini} the term $\f{\L}{8\pi G} V_3$ and treating $\L$ as a Lagrange multiplier.
The equation of motion for $\phi$ is unaffected by such term, while variation with respect to $\L$ imposes the constraint \eqref{constr}.
Concerning the $R^2$ term, we can avoid its unboundedness by imposing the kinematic constraint
\be \label{epsconstr}
\phi(t)\geq \eps \, , \;\;\; \forall t \, .
\ee
In this way we also mimic the analogous constraint that is imposed in CDT simulations.

We emphasise that the action \eqref{Smini}, except for the irrelevant topological term, reduces for $\g=0$ to the same form as \eqref{S_2d} (plus Lagrange multiplier), as we have discussed in Sec.~\ref{Sec:3d_troubles}. The term proportional to $\g$ is thus the next order correction that we were looking for.

The continuum model defines an associated BIB model with reduced transfer matrix
\be \label{BIB3d}
g(m_i,m_{i+1}) = \exp\left( -b_1 \f{2 (m_{i+1}-m_i)^2}{m_i + m_{i+1}} + b_2 \f{2}{m_i + m_{i+1}} \right)\, .
\ee
We will come back to the relation between continuous and discrete variables in the next section.  Here, we will proceed instead with the analysis in the continuum, seeking the configuration that minimises the action, thus dominating the partition function.

Before proceeding further, it is convenient to define
\be
\om^2 = \s \f{N^2 \L}{2\l-1}\, , \,\,\, \xi = \s \f{2 N^2  \g}{2\l-1}\, , \,\,\, b' = \s \f{N^2 b}{2\l-1}\, , \,\,\, \k^2=\s  \f{N G}{2\l-1} \, ,
\ee
in terms of which \eqref{Smini} reads
\be \label{Smini2}
\bar S_{\rm (2+1)-mini} = \f{ 1}{2 \k^2} \int_{-\f{\t}{2}}^{\f{\t}{2}} dt \,\left\{    \dot{\phi}^2 - \om^2 \phi^2 +b' -\f{\xi}{\phi^2}  \right\} \, ,
\ee
whose associated equation of motion is
\be \label{eom-iso}
\ddot \phi + \om^2 \phi - \f{\xi}{\phi^3} = 0\, ,
\ee
The latter is a particular case of the Pinney-Ermakov equation \cite{Pinney}, and it is exactly solvable and known to lead
to periodic solutions with the period $\t_0=\f{\pi}{\om}$, independent of $\xi$ and thus identical to the harmonic oscillator case.\footnote{\label{calogero-foot}
Interestingly, if we interpret  \eqref{Smini2} as in Lorentzian signature ($\s=-1$ and $\l<1/2$), then the associated quantum Hamiltonian is
\benn
\hat H = -\f12 \f{d^2}{d\phi^2} + \f12 \om^2 \phi^2 + \f12 \f{\xi}{\phi^2}\, ,
\eenn
which is known as the isotonic oscillator \cite{Weissman} or the one-dimensional Calogero Hamiltonian \cite{Calogero}, and which was also found in generalised models of CDT in 1+1 dimensions \cite{DiFrancesco:2000nn} (see also \cite{Ambjorn:2013hii}
for related work).
This Hamiltonian can be exactly diagonalised, and one can then compute partition function and correlators.
However, if we want to make contact with the Euclidean model, we need to analytically continue $\om\to i \,\om$ and $\xi\to-\xi$. 
The sign of $\xi$ is not necessarily a problem, as for $|\xi |<1/4$ ``fall to the center'' is avoided \cite{Calogero}. The minus sign for $\om^2$ indicates instead an unstable potential, but as we have already stressed, such a potential can only be used together with the constraint deriving from variation of the Lagrange multiplier $\om^2$. It will, hopefully, be useful to exploit this fact in future work.}
Its solution can be written as \cite{isotonic}
\be \label{phisol}
\phi_0(t) = \f{1}{\om A} \sqrt{ (\om^2 A^4 - \xi) \cos^2(\om t + \psi) + \xi } \, ,
\ee
where $A$ and $\psi$ are integration constants. By a shift of the time variable we can set $\psi=0$, so that $t=0$ corresponds to the maximum of the curve, while $A$ is fixed by initial conditions, in particular $\phi(0)=A$.
Finally, the constraint \eqref{constr} will effectively fix $\om$ as a function of $V_3$, $A$ and $\xi$.
The solution so obtained defines a universe whose spatial slices never reach zero if $\xi>0$.
As a consequence the conical singularity found in \cite{Benedetti:2009ge} is avoided, the singularity becomes a throat, or bounce, and the $S^1\times S^2$ topology can be preserved.
Furthermore, we notice that the constant solution $\phi(t) =\xi^{1/4}/\sqrt{\om}\equiv \bar\phi_0$ is a special case of \eqref{phisol} with $A=\xi^{1/4}/\sqrt{\om}$, and that $\bar\phi_0>0$ is only possible for $\xi>0$.

It is tempting to interpret the existence of the bounce and of the constant solution as indications that in the CDT model $\xi>0$. 
This could help reproduce the droplet phase along the lines of   \cite{Bogacz:2012sa}, and it could also improve it by simultaneously taking into account the non-zero spatial extension of the configurations in the stalk.
However, the constant solution to \eqref{eom-iso} cannot be joined to the oscillating one unless $A=\xi^{1/4}/\sqrt{\om}$, leading to a completely constant solution.
We will now argue that, nonetheless, a careful analysis of the minimisation of the action reveals that we can combine an oscillating solution with a constant configuration.  This configuration is not a solution to the equations of motion, but it gives an absolute (rather than local) minimum of the action. 

Let us begin by looking for a local minimum of the action, i.e. for a solution of \eqref{eom-iso}, and then proceed to evaluate the action on this configuration of $\phi(t)$.
The action to be used in this context is
\be \label{Smini3}
S_{\rm (2+1)-mini} = \f{ 1}{2 \k^2} \int_{-\f{\t}{2}}^{\f{\t}{2}} dt \,\left\{    \dot{\phi}^2 -\f{\xi}{\phi^2}  \right\} \, ,
\ee
where with respect to \eqref{Smini2} we have set $b'=0$ (the topological term only adds an irrelevant constant to the action), and we have removed the $\om^2 \phi^2$ term as this is part of the volume constraint, which vanishes on shell.
For the solution, as we said, we have two options: either it is purely oscillatory as in \eqref{phisol}, with period $\f{\pi}{\om}=\f{\t}{n}$ for some positive integer $n$, or it is a constant one $\phi(t) =\xi^{1/4}/\sqrt{\om}$.
In the first instance we find 
\be
S_{\rm (2+1)-mini}[\phi_0(t)] = \f{ n\, \pi}{4 \k^2} \left( \f{n\, \pi A^2 }{\t} -4 \sqrt{\xi} +\f{\t\xi}{n\, \pi A^2} \right)\, .
\ee
However, we still have to enforce the volume constraint \eqref{constr}, which fixes $A=A(n,V_3/N,\t,\xi)$, leading to
\be
S_{\rm (2+1)-mini}[\phi_0(t)] = \f{ n\, \pi}{8 \k^2} \left( \f{n\, V_3 }{N \t^2} -8 \sqrt{\xi} \right)\, .
\ee
Clearly the action is minimised by $n=1$, and for $\sqrt{\xi}<\f{V_3 }{8 N \t^2}$ it is positive.

In the case of constant profile, the volume constraint fixes $\om = 4\pi N\t\sqrt{\xi}/V_3$ and we find
\be \label{S_barphi0}
S_{\rm (2+1)-mini}[\bar\phi_0] = -\f{ 2 \pi N \t^2 \xi}{\k^2 V_3} \, ,
\ee
which is always negative. More importantly $S_{\rm (2+1)-mini}[\bar\phi_0]\leq S_{\rm (2+1)-mini}[\phi_0(t);n=1]$, with the equality holding only for $\sqrt{\xi}=\f{V_3 }{4 N \t^2}$, when $\phi_0(t)=\bar\phi_0$.

Thus we have found that the constant solution is favoured over the oscillating ones.
This might have been expected from  \eqref{Smini3}, where the $\dot{\phi}^2 $ term is always positive, unless $\phi$ is constant, while the remaining term is always negative for $\xi>0$.
By this argument we see that, given \eqref{epsconstr}, if it was not for the volume constraint the action \eqref{Smini3} would be minimised by $\phi(t)=\eps$. 
Notice that the latter is not a solution of the equations of motion. The fact that the partition function might be dominated by a configuration which is not an extremum of the action might look unfamiliar to the reader at first, but it is a simple consequence of the negative sign in the potential in combination with the constraint \eqref{epsconstr}, as simple examples can illustrate.\footnote{A trivial example is provided by the Gaussian integral, $I=\int_{-L}^{+L} dx\, e^{-\alpha x^2}$. For $\alpha$ very large and positive the integral is dominated by the minimum of the action at $x=0$ (of course in this case the fluctuations become essential in order to recover the actual dependence on $\alpha$, $I\sim \a^{-1/2}$); on the other hand, for $\alpha$ very large and negative the integral is dominated by the configurations $x=\pm L$ (giving $I\sim - e^{-\a L^2}/\a L$), which are not extrema of the action. Similar but more elaborate examples (e.g.~including extrema which are local minima, rather than just a local maximum as in the $\a<0$ Gaussian example) can easily be worked out.}

We see therefore that the volume constraint prevents the action from being dominated by configurations that saturate the bound \eqref{epsconstr}.
Nonetheless, the effect of the would-be unstable potential cannot be neglected, and the dominant configuration is to be expected to result from a balance of such would-be instability and the volume constraint. In particular we can expect to find (within a certain range of parameters to be determined below) a dominant configuration in which we have a stalk of minimal spatial volume occupying the largest time interval possible, combined with a short region of non-minimal spatial extension which allows the volume constraint to be respected.
In BIB models with  $g(m_i,m_{i+1}) =g(m_i)$ this spatially extended region is concentrated at a single time, which thus contains most of the total volume. In contrast to this, in models with a genuine (i.e. not ultra-local) kinetic term, such configurations would have a very large kinetic term, and thus they are disfavoured. 
Therefore we expect the spatially extended region to have a non-minimal time extension, and to be joined to the stalk in a non-singular way, e.g. by means of a $C^1$ profile.\footnote{We should stress again that the precise level of smoothness ($C^1$ versus $C^2$ or even $C^\infty$) is not important for our analysis, as long as the difference affects only a small time interval $\D t$, over which $\phi(t)$ and $\dot{\phi}(t)$ remain finite.}
In other words, we expect a time-varying droplet continuously (and at least with continuous derivative) connected to a constant stalk. The time varying part should also contribute minimally to the action, and therefore it should be a solution of \eqref{eom-iso}, while for the constant part this is not needed, as we argued.
We will therefore assume that the minimising configuration is a combination of an oscillating solution \eqref{phisol} with a constant configuration, the latter necessarily \textit{not} solving the equations of motion, i.e.
\be \label{fullsol}
\bar\phi(t) = \begin{cases} \f{1}{\om A} \sqrt{ (\om^2 A^4 - \xi) \cos^2(\om t ) + \xi } \, ,  &  \text{for } t \in [-\f{\pi}{2\om},+\f{\pi}{2\om}]\, ,  \\   \f{\sqrt{ \xi}}{\om A}  \, ,  &  \text{for } t \in [-\f{\t}{2},-\f{\pi}{2\om})\cup (+\f{\pi}{2\om},+\f{\t}{2}]\, ,  \end{cases}
\ee
with $ A^4 > \xi/\om^2$ and $\t>\pi/\om$.

In order to test the ansatz \eqref{fullsol}, we should first of all impose the volume constraint.
Plugging \eqref{fullsol} into \eqref{constr} we find
\be \label{V3constr}
V_3 = 4\pi N \left(  \f{ \pi (\xi+A^4\om^2) }{2 A^2 \om^3} +\f{\xi}{A^2\om^2} \left(\t-\f{\pi}{\om}\right)\right)\, .
\ee
The equation to be solved is cubic in $\om$, making the analysis very cumbersome, therefore it is convenient to solve it in an expansion in $\xi$, or more precisely $\xi/(A^2\om^2)$. Assuming the latter to be small implies that most of the volume is given by the droplet, with only minimal contribution from the stalk. At leading order we get
\be
V_3 = 2\pi^2  N \f{  A^2 }{ \om}\, ,
\ee
which is trivially solved for
\be \label{omegasol}
\om(A) =2\pi^2 N  \f{  A^2 }{ V_3}\, .
\ee
Now we can plug \eqref{fullsol}, with $\omega$ given by \eqref{omegasol}, into \eqref{Smini3}, to find
\be
S_{\rm (2+1)-mini}[\bar\phi(t)] = \f{1}{\k^2} \left( -\f{2 \pi^4 N^2 \t}{V_3^2} A^6 + \f{3 \pi^3 N}{2 V_3} A^4 -\pi\sqrt{\xi} +\f{V_3 \xi}{8\pi N} \f{1}{A^4}  \right)\, .
\ee
Viewed as a function of $A$ we see that it is unbounded from below, $S_{\rm (2+1)-mini}[\bar\phi(t)] \propto - A^6$, and thus minimisation favours large $A$, eventually corresponding to a delta function profile for $A\to\infty$. However, this is a degeneracy we had foreseen from the unboundedness of the action for $\xi>0$. The aforementioned cure is to impose \eqref{epsconstr}, which for \eqref{fullsol} (using \eqref{omegasol}) means
\be \label{Aeps}
A^3 \leq \f{V_3 \sqrt{\xi} }{2\pi^2 N \eps} \equiv A_\eps^3\, .
\ee
With this we obtain
\be \label{S_barphi}
S_{\rm (2+1)-mini}[\bar\phi(t);A=A_\eps] = \f{1}{\k^2} \left( -\f{\xi \t}{2 \eps^2} + \f{3}{4} \left(\f{\pi V_3 \xi^2}{2 N\eps^4}\right)^{\f13}  -\pi\sqrt{\xi} +\f12 \left(\f{\pi^5 N \eps^4 \xi}{4 V_3}\right)^{\f13} \right)\, .
\ee
We find that \eqref{S_barphi} is smaller than  \eqref{S_barphi0} for $\eps^2 \ll V_3/\t$, and hence for large volume and small $\eps$ the droplet configuration \eqref{fullsol} dominates.\footnote{Under the same conditions, our assumption that  most of the volume is given by the droplet is valid, and this proves the consistency of our analysis.
Similarly, $\t>\pi/\om$ reduces to $\t>(\f{\pi V_3}{2\xi})^{1/3} \eps^{2/3}$, which is again satisfied for small enough $\eps$.}

Defining
\be \label{bar-om}
\bar\om \equiv \om(A_\eps) = \left( \f{2\pi^2 N \xi}{V_3 \eps^2} \right)^{\f13}\, ,
\ee
we can rewrite the configuration that minimises the action as
\be \label{fullsol2}
\bar\phi(t) = \begin{cases} \sqrt{ \left(  \f{V_3 \bar\om}{2 \pi^2 N}   -\eps^2 \right) \cos^2\left(\bar\om t\right) +\eps^2 } \, ,
  &  \;\text{for } t \in [-\f{\pi}{2\bar\om},+\f{\pi}{2\bar\om}]\, ,  \\
  \eps  \, ,  & \; \text{for } t \in [-\f{\t}{2},-\f{\pi}{2\bar\om})\cup (+\f{\pi}{2\bar\om},+\f{\t}{2}]
  \, .  \end{cases}
\ee
In the limit $\xi\to 0$ with the ratio $\xi/\eps^2 =\s^2$ fixed, we recover a ``$\cos^2$'' configuration, 
\be \label{fullsol0}
\bar\phi^2(t) \to \begin{cases}   \f{V_3 \bar\om}{2 \pi^2 N}   \cos^2\left(\bar\om t\right) \, ,
  &  \;\text{for } t \in [-\f{\pi}{2\bar\om},+\f{\pi}{2\bar\om}]\, ,  \\
  0  \, ,  & \; \text{for } t \in [-\f{\t}{2},-\f{\pi}{2\bar\om})\cup (+\f{\pi}{2\bar\om},+\f{\t}{2}]
  \, ,  \end{cases}
\ee
while preserving the condition $\pi/\bar\om <\t$. We should notice however that this is not a 3-sphere yet.
Introducing the following notation,
\be \label{rugby-vol}
V_3 = 2\pi^2  s\, r^3 \, , \;\;\; s= \f{N}{\s}
\ee
and using it in \eqref{fullsol0}, we arrive at $\bar\phi^2(t) =  r^2 \cos^2( \f{N t}{ r s})$, for the extended part of the universe.
Changing time variable to $\psi= \f{N t}{ r s}$, we obtain the following line element for the spacetime metric
\be \label{rugby-ds}
ds^2 = r^2 (s^2 d\psi^2 +  \cos^2(\psi) d\Om_2  ) \, , 
\ee
where $d\Om_2$ is the standard line element on the 2-sphere. 
We recognise in \eqref{rugby-ds} the line element of the stretched/squashed 3-sphere discussed in \cite{Benedetti:2009ge}. The latter has conical singularities for $s\neq 1$ at $\psi=\pm\pi/2$, while as we noticed before, for $\xi\neq 0$ such singularity is lifted in \eqref{fullsol2}.

Due to the approximation used in obtaining \eqref{omegasol}, the configuration \eqref{fullsol2} violates \eqref{V3constr} by terms of order $\eps^2$. In fact  \eqref{omegasol} was obtained from an expansion in $\xi/(A^2\om^2)$, which by \eqref{Aeps} is turned precisely into $\eps^2$.
Improving the expansion amounts to replacing everywhere in  \eqref{fullsol2} $V_3\to\Vt_3$, with $\Vt_3$ such that  \eqref{V3constr} is satisfied. We find
\be \label{tildeV3}
\Vt_3=V_3 - 4\pi N \eps^2 \t + O(\eps^{8/3}) \, ,
\ee
or, keeping $\xi/\eps^2 =\s^2$ fixed,
\be \label{tildeV3-bis}
\Vt_3=V_3 - 4\pi N \eps^2 \t + \left(2\pi^2\right)^{\f23} \left(\f{V_3}{\s^2}\right)^{\f13} \eps^2 + O(\eps^4) \, .
\ee

In this way, we find a $\t$-dependence in $\omega$.
This is described by $\eps^2 \t$ corrections to \eqref{omegasol}, and hence constant $\om$ is a good approximation for small $\eps$ and values of $\t$ that are not too large.
In any case, our CDT simulations have been carried out at a single value of $T$, so we cannot use them to test the $\t$-dependence of $\om$ in detail. 
However, some results on the $T$-dependence of the simulations have been discussed elsewhere in the literature, and we can compare them to some qualitative features of our analysis.
Simulations at different values of $T$ have been presented in \cite{Cooperman:2013mma}, and from their analysis it can be read off that the width of the droplet is roughly insensitive to $T$ as long as $T$ stays larger than the width of the droplet (i.e. larger than $\a N_3^{1/3}$ for some critical $\a$). 
In \cite{Ambjorn:2000dja} it was also reported that below a certain critical value of $T$ a uniform distribution of spatial volumes was observed. 
Such observations are consistent with our results, because if $\t$ becomes sufficiently smaller than $V_3^{1/3}$ then the second term in \eqref{S_barphi} will become the dominant contribution, making the action positive, and thus larger than  \eqref{S_barphi0}.
This happens roughly at the same value for which the condition $\pi/\bar\om <\t$ breaks down (in this case, again the contest is between \eqref{phisol} and $\bar\phi_0$, and we have already shown that the latter wins), so we can write
\be
\t_- \simeq  \left( \f{\pi V_3 \eps^2}{2 N \xi} \right)^{\f13} \, ,
\ee
for the critical value of $\t$ below which (in the leading order solution \eqref{omegasol}) the constant solution becomes the dominant configuration.
We have also checked numerically that solving the full constraint \eqref{V3constr} for different values of the parameters, the picture is fully consistent with the approximate solution above.

From our analysis, because of the quadratic dependence on $\t$ in \eqref{S_barphi0}, versus the linear one in  \eqref{S_barphi},  we can also predict that for $\t$ larger than some critical value $\t_+$, the constant solution will dominate again. 
However, there is a limit to how large $\t$ can become, since for $\t>\t_{\rm max}\equiv V_3/(4\pi N \eps^2)$ the kinematical constraint $\phi(t)>\eps$ becomes incompatible with the volume constraint.
Coincidentally, $\t_{\rm max}$ is also the value at which the leading term in  \eqref{S_barphi} becomes equal to  \eqref{S_barphi0}.
For small $\eps$, the second term in \eqref{S_barphi} dominates over the last two, and since it is positive, we can deduce that $\t_+<\t_{\rm max}$.

To summarise, we conclude that for $\t_-< \t < \t_+$ the partition function for the minisuperspace model defined by the action \eqref{Smini3}, together with the volume constraint \eqref{constr} and the kinematical constraint \eqref{epsconstr}, is dominated by the configuration \eqref{fullsol2}, where  $V_3\to\Vt_3$.

\section{Comparison between the HL-minisuperspace model and CDT}
\label{Sec:comparison}

In the previous section, we have shown that the minisuperspace model inspired by HL gravity is successful in qualitatively reproducing the droplet condensation of (2+1)-dimensional CDT, thus solving the problem that we encountered when attempting the same analysis with its GR counterpart ($\g=0$).
It should be stressed that in achieving this result, an important role was played by the presence of the constraint  \eqref{epsconstr}.
As mentioned, this is reminiscent of the constraint typically imposed on the CDT configurations, and so it fits nicely into the comparison.
However, in CDT the constraint has always been viewed up to now as a discretisation artifact, a view supported by the fact that the extension of the universe within the stalk goes to zero in the continuum limit (see Fig.~\ref{Fig:scaling}). Therefore one might be suspicious that we are attributing too much importance to a ``lattice artifact''.  However, the two statements (that $\eps>0$ plays an important role, and that $\eps\to 0$ in the continuum) are perfectly consistent with each other, as we will now explain.

As we discussed in Sec.~\ref{Sec:data}, without any fine tuning, quantities of positive length dimension tend to go to zero with the cutoff. The obvious exception is the total volume, which we are increasing by hand. This is equivalent to fine tuning the cosmological constant to its critical value. Simulations show that the shape of the droplet scales with the volume  (Fig.~\ref{Fig:scaling}); in other words we naturally have a continuum limit with fixed $A$ and $\om$. However, Fig.~\ref{Fig:scaling} shows also that the total time extension and the spatial extension of the stalk both go to zero.
In other words, $\t$ and $\eps$ go to zero. This is exactly what dimensional analysis leads us to expect: $\t$ and $\eps$ have both dimension of length, so without any fine tuning they naturally scale like the cutoff $a$.
The important point is that the same fact holds true also for $\xi$, which has dimensions of (length)${}^2$. This being so, we expect that in the continuum limit the ratio $\s^2=\xi/\eps^2$ stays roughly constant (in fact we have that $A_\eps \bar\om/\pi=\s/\pi$ is the ratio between the amplitude and the width of the droplet, which  Fig.~\ref{Fig:scaling} shows to be a constant), and as a consequence we should obtain the configuration \eqref{fullsol0} as a final result, while if started with $\xi=\eps=0$ from the beginning this would not be possible.

It is therefore interesting to ask if we can fine tune the model in order to keep all the dimensionful couplings in \eqref{fullsol2} finite.
Concerning $\t$, we can trivially avoid its shrinking, as a scaling of $T$ can be introduced by hand, just as is done for the total volume.  We could re-do the simulation for various values of $N_3$ with a constant ratio $T/N_3^{1/3}$. We expect that this will not change any features of the results, except preserving the ratio between the time extensions of stalk and blob.

We could also scale $n_\eps\sim \beta N_3^{2/3}$ in \eqref{minN2} in order to have a finite $\eps$ in the continuum limit  (keeping $\b\ll 1$ so that the stalk remains smaller than the blob)\footnote{To be precise, the spatial volume within the stalk is slightly larger than the minimal allowed value (while $n_\eps=4$, we have $n_s\sim 10$). This is of course expected: a distribution of numbers greater or equal to $n_\eps$ cannot average to $n_\eps$ unless they are all precisely equal to it. However, since the difference between $n_\eps$ and $n_s$ does not scale with the total volume, we expect to be able to control $n_s$ by varying $n_\eps$.}.
 In contrast, we do not have direct control over $\xi$,\footnote{See \cite{Anderson:2011bj} for simulations of a CDT model with a discrete version of the spatial $R^2$ term in the action.}
 so we do not know whether scaling $n_\eps$ would be sufficient in order to keep a finite $\xi$ as well.
In the positive case, by repeating a plot like  Fig.~\ref{Fig:scaling} we should see a superposition of data both for the blob and for the stalk, and we would interpret $\xi$ as being associated with the kinematical constraint $N_2(i)\geq n_\eps$.
In the negative case, in the same type of plot we should lose the superposition even in the blob, and we would conclude that $\xi$ is an independent and intrinsic feature of the model, which needs to be tuned separately.  We hope to be able to perform such test in the near future.

We conclude our work by performing a quantitative comparison between our theoretical model and the CDT data.
To that end we need to rewrite \eqref{fullsol2} in discrete variables.  A careful presentation of such ideas can be found in \cite{Cooperman:2013mma}, which we will follow with a few variations.  First of all, we define $\Nt_3$ as the discrete volume being held fixed, which in principle is $N_3$, but as we saw (Fig.~\ref{Fig:N_2-tot}), to a good approximations can also be taken to be $\la N_2^{(s)} \ra = \sum_i \la N_2(i) \ra = \f12 \la N_{(3,1)} \ra$. 
Then we take the relations
\be
V_3 = v_3\, a^3 \, \Nt_3\, ,
\ee
\be
N\, t = \a \left(\f{V_3}{\Nt_3}\right)^{\f13} i\, ,
\ee
and
\be
\int dt \, N \, V_2(t) = v_3\, a^3 \sum_i \la N_2(i) \ra\, ,
\ee
from which, stripping off integral and sum in the latter, we obtain
\be
\la N_2(i) \ra = \f{ dt \, N }{v_3 \, a^3} \, V_2(t)   = \a \left(\f{\Nt_3}{V_3}\right)^{\f23}  \, V_2\left( \f{\a}{N} \left(\f{V_3}{\Nt_3}\right)^{\f13} i \right)  \, . 
\ee
Using \eqref{fullsol2} in \eqref{V2}, with the substitutions in \eqref{bar-om} and \eqref{tildeV3-bis}, we obtain
\be \label{fullsol-discr}
\la N_2(i) \ra = \begin{cases} \left( \f{2}{\pi}  \f{\r^2 \Nt_3^{2/3}}{\chi} - n_s\right) \cos^2\left(  \f{ i}{\r\, \chi \Nt_3^{1/3}} \right) +n_s \, ,
  &  \;\text{for } i \in [-\f{\pi}{2} \r\, \chi \Nt_3^{1/3},+\f{\pi}{2} \r\, \chi \Nt_3^{1/3}]\, ,  \\
  n_s  \, ,  & \; \text{else } 
  \, , \end{cases}
\ee
where we have identified
\be
n_s = \f{4\pi \eps^2 N \t}{v_3 a^3 T} =  4\pi \eps^2 \a\left(\f{\Nt_3}{V_3}\right)^{\f23} \, ,
\ee
and we have defined
\be
\chi = \f{s^{2/3}}{\a (2\pi^2)^{1/3}} \, , \;\;\;\;\; \r = \left(\f{\Vt_3}{V_3}\right)^{\f13} \, .
\ee
The latter are two independent parameters, in particular $\r$ depends explicitly on $\eps$ (see \eqref{tildeV3-bis}). However, we expect $\r\to 1$ in the continuum limit.

\begin{figure}[ht]
\centering 
\includegraphics[width=13.5cm]{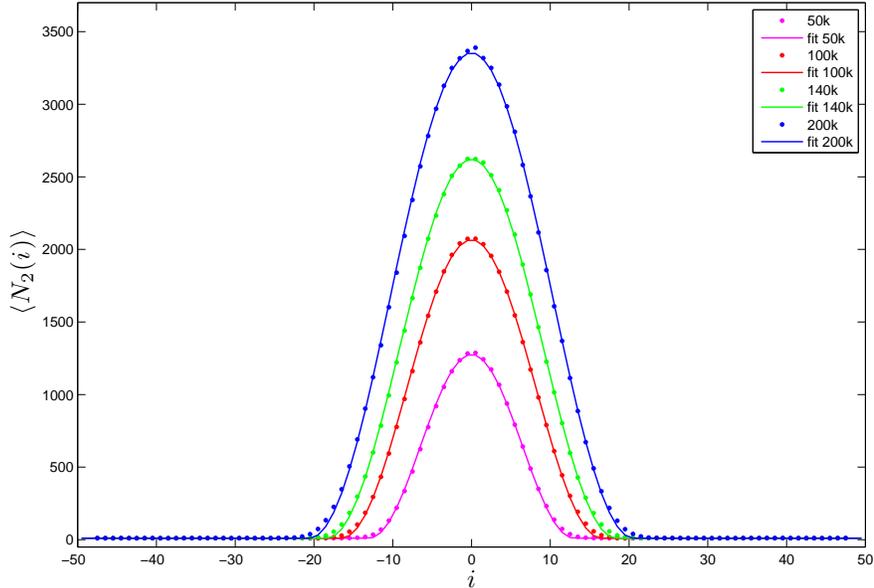}
\caption{\small{Plots of $\la N_2(i) \ra$ for the CDT model at $\kappa_0=5$, together with the results of fits based on \eqref{fullsol-discr}. The error bars are typically of the size of the markers, or smaller, and thus are omitted. The only exception is for the few points around the peak of the droplet in the 200k case, for which errors are about double the size of the markers. This is due to the smaller number of Monte Carlo configurations in that case.}}
\label{Fig:fits}
\end{figure}

For $n_s=0$ (i.e. $\eps=0$, hence $\r=1$), \eqref{fullsol-discr} is exactly the same type of function used in \cite{Cooperman:2013mma}\footnote{Their $\tilde s_0$ corresponding to our $\chi$ for fitting purposes, although their definition and interpretation is slightly different.} (and is the analogue of the function used in 3+1 dimensions in \cite{Ambjorn:2007jv,Ambjorn:2008wc}).
The explicit presence of $n_s\neq 0$, and hence also the presence of $\r$, in our function is thus the main practical difference for fitting purposes.\footnote{As a matter of fact, the stalk is taken into consideration by Cooperman and Miller in \cite{Cooperman:2013mma}.  However, they insisted on a purely $n_s=0$ ansatz for the blob, but $n_s\neq 0$ for the stalk. As a consequence, they have to join the two functions at $|i|<\f{\pi}{2} s_0 \Nt_3^{1/3}$, leading to a discontinuity in the first derivative of the profile function.}
However, given the smallness of $n_s/\Nt_3^{2/3}$, we do not expect significant differences between our fit and previous ones.

A comparison between fits and data is shown in Fig.~\ref{Fig:fits}, and the numerical values for the fits are given Table~\ref{table}.
The fitting parameters turn out to be only mildly dependent of the system size. 
In the last column of the table we give values for $s$ in the case that $\a=1$. We obtain in general $s<1$, meaning that we have a squashed sphere, which is consistent with inspection of the volume profiles (in particular the rescaled ones in Fig.~\ref{Fig:scaling}), but we should stress that there is so far no reason to prefer $\a=1$ with respect to other values.
In the last row of the table we give the results of linear extrapolations on the fitting parameters. These, however, should be taken with a grain of salt, as we are fitting only four points, and moreover, with the exception of $\r$, the linear fit fares very poorly. The relevant message from such extrapolations is that we find that $\r$ is approaching unity as $N_3$ grows. This is consistent with our expectations, because $\r=1$ for $\eps=0$, and due to the fact that $\eps$ has the dimension of a length, we expect $\eps\to 0$ in the continuum limit. Such expectation is substantiated by Fig.~\ref{Fig:scaling}, where one notices that the volume in the stalk scales to zero. The extrapolation for $\r$ in Table~\ref{table} provides a quantitative translation for such a qualitative observation.

\begin{table}
\begin{center}
\begin{tabular}{|c|c|c|c|c|c|}\hline
$N_3$  & $\tilde N_3 = \la N_2^{(s)} \ra$ & $n_s$ & $\r$ & $\chi$ & $s$ \\ \hline
50k \;&\;  17548 \;&\; 10.60 \;&\; 0.9809 \;&\; 0.3239  \;&\; 0.8191 \\ \hline
100k \;&\;  35026  \;&\; 10.57 \;&\; 0.9903 \;&\; 0.3233  \;&\; 0.8168 \\ \hline
140k \;&\; 49010 \;&\; 10.55 \;&\; 0.9930 \;&\; 0.3206  \;&\; 0.8066 \\ \hline
200k \;&\; 69983 \;&\; 10.40 \;&\; 0.9949 \;&\; 0.3190  \;&\; 0.8004 \\ \hline\hline
$\infty$ \;&\; $\infty$ \;&\; 10.43 \;&\; 0.9997 \;&\; 0.3186  \;&\; 0.7989 \\ \hline
\end{tabular}\end{center}
\caption{Fitting parameters for Fig.~\ref{Fig:fits}. In the actual fit we have only three parameters, $n_s$, $\chi$ and $\r \Nt_3^{1/3}$. We use the values of $\Nt_3 = \la N_2^{(s)} \ra$ to extract $\r$, and we assume $\a=1$ to extract $s$. The last row is a linear extrapolation for the fitting parameters versus $1/\Nt_3$.}
\label{table}
\end{table}

We notice that the fit for the 200k data is not very good near the peak, however, the data themselves are not particularly good (there is clearly one point stemming out of the group), having in particular larger errors near the peak of the droplet, which we attribute to the poorer statistics at this system size. It should be mentioned that to the best of our knowledge, this is the largest system size ever reported for (2+1)-dimensional CDT (\cite{Ambjorn:2000dja} reached 64k at most, and \cite{Cooperman:2013mma} reached 102k), and so we hope that we shall be forgiven for not pushing the simulations even further.

It should also be mentioned that in \cite{Cooperman:2013mma} Cooperman and Miller noticed that fits perform better for certain values of $T$, and so it is possible that the 200k simulations happen to have an unlucky ratio between blob width and total time extension $T$.

Lastly, we point out that in \cite{Bogacz:2012sa}, Bogacz et al. found that a better quantitative agreement with direct Monte Carlo simulations of the BIB model is obtained by including the effect of quadratic fluctuations (i.e. computing the one-loop effective action) before minimising over $A$. This might also be relevant for CDT. We will return to this question in future work.

\section{Conclusions}
\label{Sec:concl}

We have argued that in order to explain the spacetime condensation phenomenon of CDT presented in Sec.~\ref{Sec:data}, a minisuperspace model based on GR is not sufficient.
By finding and studying a minisuperspace model inspired by HL gravity, we have shown that a semiclassical analysis leads to a condensation compatible with the one observed in CDT.
In doing that we have shown that higher order terms in the spatial curvature are necessary both in reproducing the droplet phase and in curing the conical singularities we encountered in \cite{Benedetti:2009ge}.
An important role was also played by the constraint $\phi(t)\geq \eps > 0$, which mimics the similar constraint imposed on the triangulations. In the continuum we can have $\xi\sim\eps^2\to 0$, but at finite cutoff the effective action should be of the HL-type in order to obtain a droplet. 
The main point we wish to highlight is that for $\xi=0$ we cannot recover a droplet, while this is possible with $\xi\neq 0$. 
We should also point out that the presence (and sign) of such term most likely depends on the spatial topology. In fact, from \cite{Budd:2013waa} we learn that CDT in 2+1 dimensions with the spatial topology of a torus and periodic time does not show signs of condensation.\footnote{We would like to thank Renate Loll for making us aware of this result.}
This can be understood from our HL-inspired model by reviewing the way in which the action \eqref{actionHL} reduced to \eqref{Smini}. If, in that derivation, we replace the auxiliary metric $\hat g_{ij}$ with the standard metric on the torus, we find that the associated scalar curvature $\hat R$ is zero, and therefore we have no potential term in the minisuperspace action. In that case we already know that that the dominant configuration is a constant volume profile, in agreement with the results of \cite{Budd:2013waa}. It would be interesting to test (2+1)-dimensional CDT with spatial slices of higher genus, for which our analysis would again suggest a condensation.

One might wonder whether or not the action we have proposed is unique. We do not make any claim of uniqueness of the action (in fact we do not expect it to be the end of the story), and higher order terms in the curvature might be necessary in order to explain small effects that are visible only at larger system size. Such terms would probably not affect the large scale properties discussed here, and thus adding them would be likely to yield a large class of actions that lead to condensation.\footnote{We emphasise once more that in the present case the effect of an $R^2$ term was important only because in 2+1 dimensions the linear $R$ term is of topological nature.}
However, the important point is that none of them would be a discretisation of the (minisuperspace) GR action. Furthermore, given that in this case we have no reasons based on symmetry to enforce special relations between spatial curvature terms and time-derivative terms,\footnote{In fact, there is presently not much support for the presence of higher-order time derivatives in the effective action \cite{Ambjorn:2011ph}.} we conclude that possible effective actions of this type all correspond to some version of (discretised, minisuperspace) HL gravity.

There are reasons to suspect that HL-type corrections to the effective action might be present also in 3+1 dimensions. Besides the similarities between HL gravity and CDT already pointed out in the literature \cite{Horava:2009if,Ambjorn:2010hu}, we stress here the importance that the higher curvature term had in our analysis in order to remove the conical singularities of the stretched/squashed sphere. As such singularities are also found in 3+1 dimensions, it is plausible to expect that they should be cured by higher order terms in the effective action, and because the stretching/squashing naturally introduces an anisotropy in the continuum, it is reasonable to expect that such higher order terms must also be of an anisotropic nature.

Many questions remain open, some of which we briefly discuss here.  We plan to return to them in a more detailed study.

First of all, the analysis presented here is of semiclassical (or mean field) type, i.e. we simply looked for a minimisation of the action.
This approximation should be put under scrutiny, comparing it to fully nonperturbative results obtained by direct numerical simulations of the associated BIB model, as well as by studying the effects of fluctuations. Direct simulations of the BIB model would also allow us to explore other phases of the model, and possibly make contact with the results of  \cite{Anderson:2011bj} on the extended phase diagram.

It is also important to perform more CDT simulations in 2+1 dimensions, for several reasons. One reason is to test the possibility of scaling $n_\eps$ (and $T$) in order to obtain a continuum limit with finite $\xi$ and $\eps$, as discussed in Sec.~\ref{Sec:comparison}. 
Another reason would be to check how all the effective couplings depend upon the (inverse) bare Newton's coupling $\k_0$. It is known (e.g. \cite{Ambjorn:2008wc}) that the period and amplitude of the blob change with $\k_0$. As Newton's constant $G$ plays no role in the classical solution (in the absence of matter), such phenomena remain unexplained in our treatment, indicating that a simple identification of $\k_0$ with the inverse effective Newton's constant might be too naive.\footnote{This is actually not a surprise, and a more careful identification was suggested in \cite{Ambjorn:2008wc} for the higher-dimensional case.} A detailed study of the $\k_0$ dependence of the shape of the CDT droplet would also allow us to set up a renormalisation group analysis along the lines of \cite{Ambjorn:2014gsa}.
Lastly, it would be very interesting to attempt to extract the effective action directly from the CDT data, as attempted in \cite{Ambjorn:2012pp,Ambjorn:2011ph} in 3+1 dimensions.

As a last remark, we have stressed above that a Hamiltonian constraint should not be imposed when studying the effective dynamics for the available CDT data. In principle, the Hamiltonian constraint should become enforced only once we integrate over the full time extension $\t$, from minus to plus infinity. However, we should point out a potential obstruction on this route: according to our analysis, $\t$ is bounded to a finite interval $\t\in [\t_-,\t_+]$, outside of which we leave the droplet phase. It is not clear to us whether the Hamiltonian constraint could be recovered in the presence of such a bound, and we hope that this will be clarified in the near future.

\subsection*{Acknowledgements}

JH receives support form EPSRC grant \textit{DIQIP} and ERC grant \textit{NLST}.


\providecommand{\href}[2]{#2}\begingroup\raggedright\endgroup


\end{document}